\documentclass[%
aip, pop,
reprint,    
superscriptaddress,
amsmath,amssymb,
aps,
]{revtex4-1}
\usepackage{graphicx}
\usepackage{dcolumn}
\usepackage{bm}
\usepackage{amsmath}
\usepackage{hyperref}
\usepackage{soul, color, xcolor}
\hypersetup{hypertex=true,colorlinks=true,linkcolor=blue,anchorcolor=blue,citecolor=blue,urlcolor=blue}
\usepackage{lineno}
\usepackage{enumerate}
\usepackage{color}
\usepackage{float}
\begin{document}

\title{Energy coupling in intense laser solid interactions: material properties of gold}

\author{X. Liu}
\affiliation{Key Laboratory for Laser Plasmas and Department of Physics and Astronomy, Shanghai Jiao Tong University, Shanghai 200240, People’s Republic of China}
\affiliation{Collaborative Innovation Center of IFSA (CICIFSA), Shanghai Jiao Tong University, Shanghai 200240, People’s Republic of China}

\author{D. Wu}
\email{dwu.phys@sjtu.edu.cn}
\affiliation{Key Laboratory for Laser Plasmas and Department of Physics and Astronomy, Shanghai Jiao Tong University, Shanghai 200240, People’s Republic of China}
\affiliation{Collaborative Innovation Center of IFSA (CICIFSA), Shanghai Jiao Tong University, Shanghai 200240, People’s Republic of China}

\author{J. Zhang}
\email{jzhang@iphy.ac.cn}
\affiliation{Key Laboratory for Laser Plasmas and Department of Physics and Astronomy, Shanghai Jiao Tong University, Shanghai 200240, People’s Republic of China}
\affiliation{Collaborative Innovation Center of IFSA (CICIFSA), Shanghai Jiao Tong University, Shanghai 200240, People’s Republic of China}
\affiliation{Institute of Physics, Chinese Academy of Sciences, Beijing 100190, People’s Republic of China}
\date{\today}

\begin{abstract}
In the double-cone ignition inertial confinement fusion scheme, high density DT fuel is rapidly heated with high-flux fast electrons, which are generated by short and intense laser pulses. Gold cone target is usually used to shorten the distance between the critical surface and the compressed high density DT core. The material properties of solid gold may affect the generation and transport of fast electrons significantly, among which the effects of ionization and collision are the main concerns. In this work, the effects of ionization, collision and blow-off plasma on laser energy absorption rate are investigated using the LAPINS code: A three-stage model is adopted to explain the mechanism of fast electron generation and the change in laser energy absorption rate. With the increase of the charge state of Au ions, the laser-plasma interaction transfers to the later stage, resulting in a decrease in laser energy absorption rate. Collision has both beneficial and harmful effects. On one hand, collision provides a thermal pressure that makes it easier for electrons to escape into the potential well in front of the target and be accelerated in the second stage. On the other hand, collision increases stopping power and suppress electron recirculation within the target in the third stage. The vacuum sheath field behind the target enhances the electron circulation inside the target and thus improves the laser energy absorption, however this effect will be suppressed when the blow-off plasma density behind the target increases or collision is considered.
\end{abstract}

\maketitle

\section{Introduction}\label{sec:first}
Inertial confinement fusion (ICF) has been proposed and studied for decades as one of the two main paths to achieve stable and controllable fusion \cite{r1,r2}. In ICF, deuterium-tritium fuel is compressed to a state of high density and high temperature under the action of the driver energy (e.g., superintense laser, ion beams or Z-pinch device), and the plasma is confined long enough by its inertial for the thermonuclear burn to produce copious amounts of fusion energy. In 1972, Nuckolls first came up with the idea of compressing tiny targets with high-power lasers to bring thermonuclear fuel to ignition conditions \cite{r3}.  Then an approach to ICF, known as the Fast Ignition scheme (FI) is proposed in which precompressed fuel is ignited by an external hot electron source \cite{r4,r5,r6,r7}. In principle, fast ignition can contribute to a higher gain than the conventional center ignition scheme. In addition, due to the separation of ignition process and the implosion process, the limits of compression symmetry and hydrodynamic instability in conventional hot-spot ignition scheme could be relaxed in FI \cite{r8}.

Double-Cone Ignition scheme (DCI) \cite{r9} is a newly-proposed method that involves four processes: quasi-isentropic compression, acceleration, collision of fuels and rapid heating. The acceleration and collision process of DCI are able to pre-heat the fuel to provide 20$\%$-30$\%$ of the energy required for ignition. As a result, the energy requirement of picosecond heating laser pulses can be significantly reduced. Through the symmetrical collision process of two high-speed fuel targets, the newly generated fuel can reach a temperature of about 1 keV and double its density. Finally in the rapid heating process, guided by an applied magnetic field, MeV electron beams generated by the interaction of picosecond heating laser pulses and several gold cones can reach the core area of the fuel, heating the fuel to ignition temperature. To improve the energy coupling and achieve a higher gain, it is of broad significance to study the role of gold cone in the generation and transport process of fast electron beams. 

The optimal geometric parameters of the cone, such as cone angle and cone structure, have been studied \cite{r10,r11,r12} and widely accepted. There are also multidimensional simulations of the interaction between laser and Au cone to study the hot-electron generation and their transport inside the Au cone \cite{r13,r14}. However, the mechanism of laser-solid gold cone interaction and how the properties of gold cone affects the generation of fast electrons are not fully understood. For the convenience of simulation, density of solid gold is artificially decreased and the binary collision between electrons and ions is often ignored in past research. In our simulations, the high electron density of the gold cone is the same as the real gold cone, and we take into account collisions between particles which is handled based on Monte Carlo method to bring our results closer to the real situation. By the way, in the simulation considering collision process, the reported results are divergent. Some researchers believe that collision will reduce the electron supply to the laser-plasma interaction (LPI) region therefore is harmful to the laser-target energy coupling \cite{r15}. On the contrary, some other researchers believe that collision helps to delay the formation of electron density steepening at the interface so it is beneficial for energy coupling \cite{r16}.

In this paper, using the recent developed PIC code LAPINS \cite{r17,r18,r19}, the coupling of multiple physical processes in laser-solid gold interaction is studied. The results show that the process of laser-plasma interaction could be divided into three different stages. Laser energy absorption rate decreases due to the transition of the laser-plasma interaction to the later stage when charge state of Au ions increases. Collision has both beneficial and harmful effects. On the one hand, collision provides a thermal pressure which makes it easier for the electrons to escape into the potential well in front of the target to be accelerated in the second stage. On the other hand, collision increases the stopping power and suppresses the electron recirculation within the target in the third stage. The vacuum sheath field behind the target enhances the electron circulation inside the target and thus improves the laser energy absorption, however this effect will be suppressed when the blow-off plasma density behind the target increases or collision is considered. Our results may provide references for the on-going DCI champaign in the gold cone target design used for rapid electron heating.

The paper is structured as follows: Sec II introduces the simulation setup parameters and some information about the LAPINS code. In Sec III, the time evolution of laser-plasma interaction process considering dynamic ionization is analyzed. A three-stage model of laser plasma interaction is given to explain how the absorption rate of laser energy is influenced by ionization and collisional effects. In order to verify the three-stage model by controlling the variables, simulations of different fixed charge state with and without collision are also carried out. In Sec IV, the relationship between laser energy absorption rate and target charge state is presented. The role of the collisional effects in laser-target energy coupling is then discussed in Sec V. And the collisional effects on electron recirculation in the presence of blow-off plasma is discussed in Sec VI. Finally, summary of this paper is given in Sec VII.

\maketitle
\section{Model description}
\begin{figure}[h]
  \includegraphics[scale=0.325]{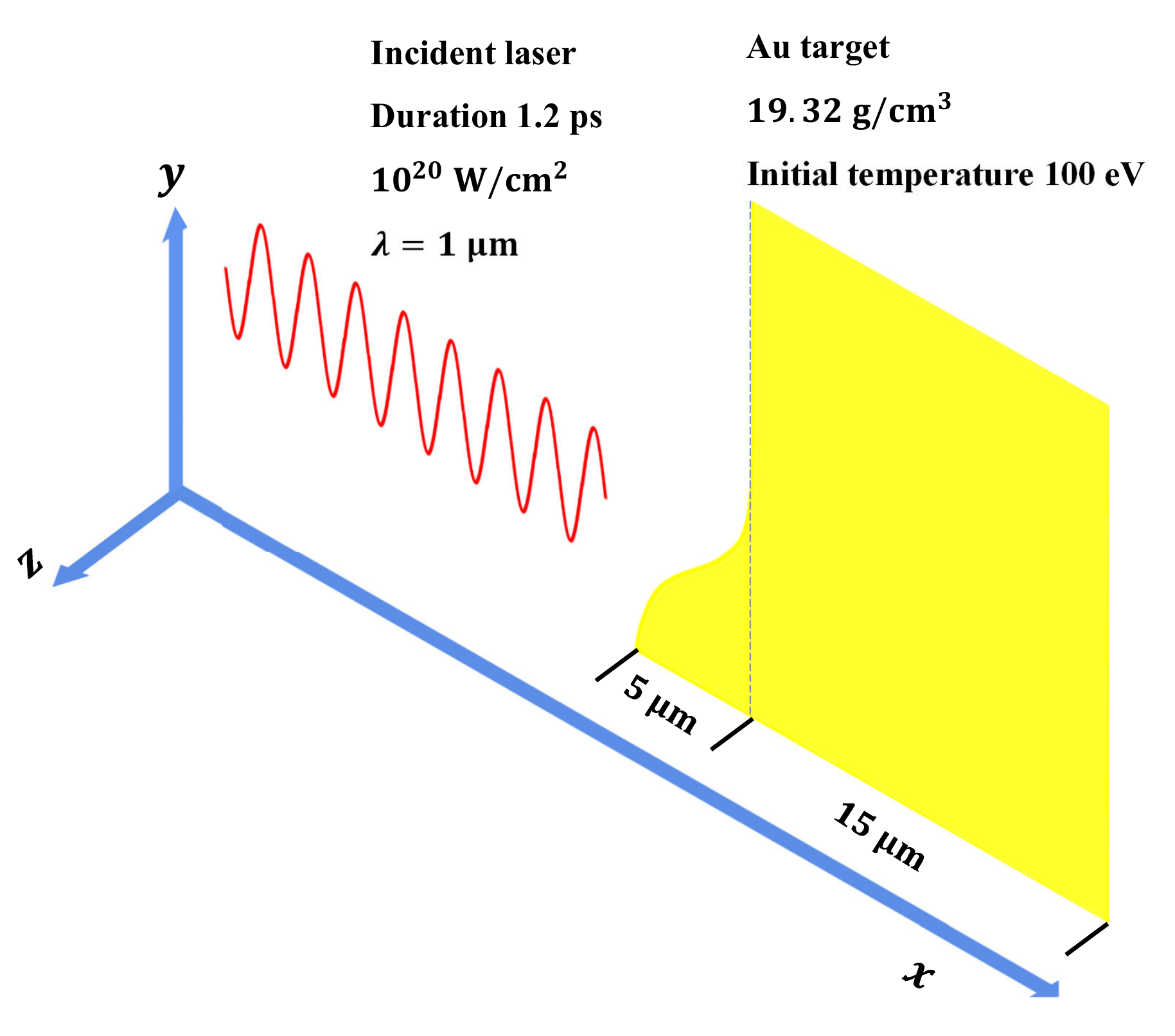}
  \caption{\label{fig:figure1} Schematic of the simulated situation.}
\end{figure}

The simulation is carried out using the PIC code LAPINS \cite{r17,r18,r19}. In LAPINS, multiple physical effects such as collision \cite{r20}, ionization \cite{r21,r22}, radiation \cite{r18}, QED \cite{r23} and nuclear reactions \cite{r24}, are included and coupled. To simulate the interactions between laser and matter with a large number of particles, the weighted particles technique is used in simulations, which has proven to be more efficient than the uniformly weighted particles in the calculation \cite{r25}. The collision model in our PIC code is based on Monte Carlo binary collisions \cite{r20}, including binary collisions among ion-electron, ion-ion and electron-electron. Contributions of both free and bound electrons are considered in the model. The calculation of the collision process is carried out in three steps: (i) pair of particles are randomly selected from the cell, which may be ion-electron, ion-ion or electron-electron pair; (ii) for the selected pair of particles, we calculate the particle velocity change due to Coulomb collision within the time interval; (iii) replace the velocity of each particle by the newly calculated one. For the ionization module, our code includes field ionization (FI), collision ionization (CI) based on the electron-ion collision cross sections, electron-ion recombination (RE) based on three body recombination, and ionization potential suppression (IPD) model \cite{r21,r22}. In addition, a high-order implicit numerical method is used in our LAPINS code to avoid numerical heating and reduce the calculation burden by using large grid size \cite{r17,r18}. The 1D simulation box is 35 $\mu$m long with grid resolution of 0.01 $\mu$m in which 1000 particles per cell are placed. The linearly polarized laser is incident from the left, with a wavelength of $\lambda=$ 1 $\mu$m and an intensity of $I=10^{20}$ W/cm$^2$. The rising time and dropping time of the incident Gaussian profile laser are 100 fs and the flat time is 1 ps. The simulation time is set to 1.2 ps, which is equal to the incidence time of the laser. The absorbing boundary condition of fields and particles are adopted in the direction of laser propagation. The 15 $\mu$m Au target is placed at the end of the simulation box, with solid density of 19.32 g/cm$^3$ and initial temperatures of 100 eV for both ions and electrons. In front of the target, 5 $\mu$m pre-plasma is attached in which electron number density increases from 3.5$\times$10$^{19}$ cm$^{-3}$ to 5$\times$10$^{21}$ cm$^{-3}$ exponentially with the scale length of 1 $\mu$m. Two observation planes are set at $x=21$ $\mu$m and $x=34$ $\mu$m respectively to count the number and energy of fast electrons passing through them. The schematic of the simulation is shown in Fig. \ref{fig:figure1}. In cases where the charge state of Au ions is calculated dynamically in the next section, the initial charge state of Au ions is $Z=5$.

\maketitle
\section{Three-stage model of laser plasma interaction}
\begin{figure}[h]
  \includegraphics[scale=0.25]{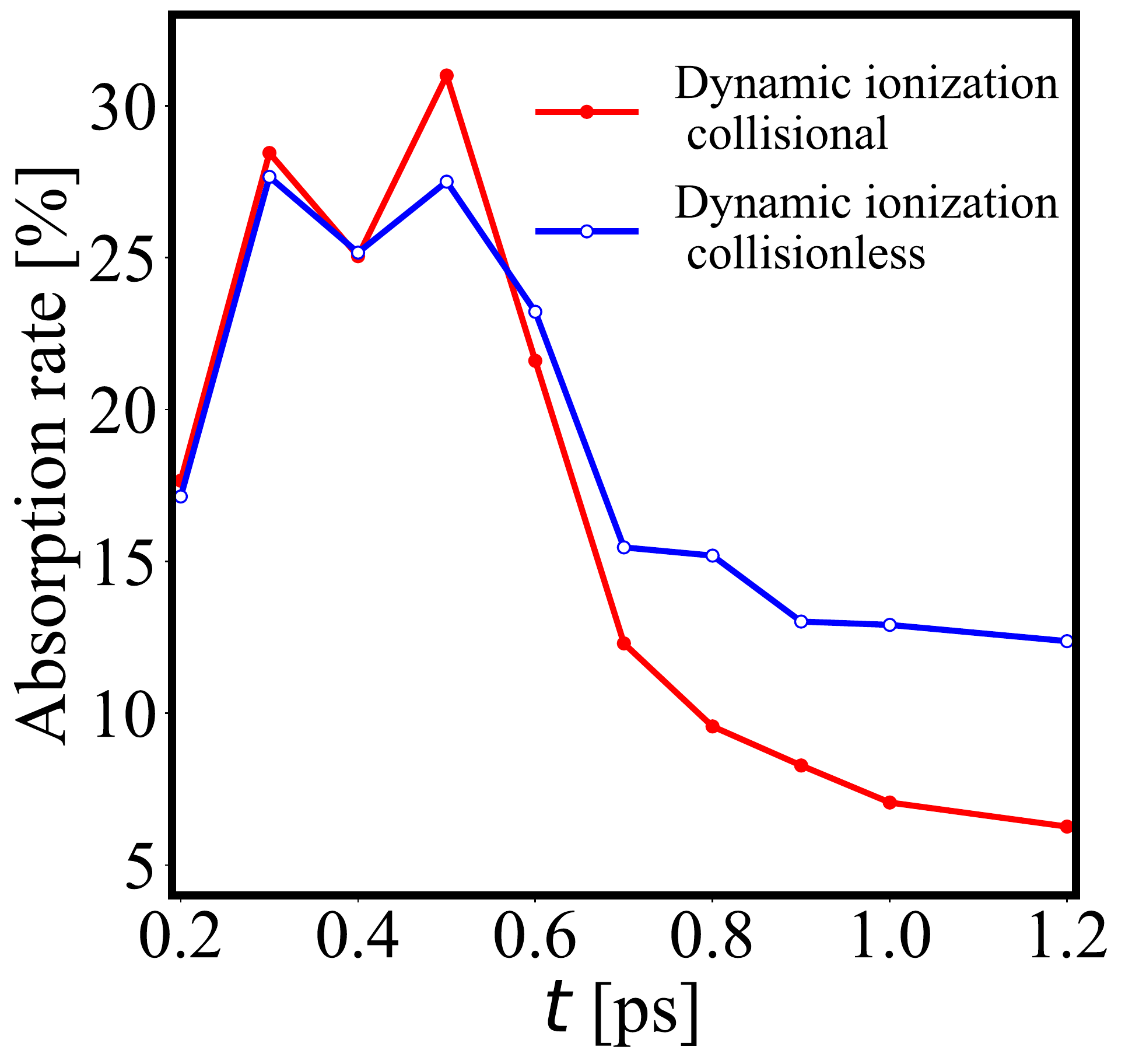}
  \caption{\label{fig:figure2} Evolution of laser energy absorption rate calculated every 0.1 ps for dynamic ionization cases. (red) Collisional case, (blue) collisionless case.}
\end{figure}

In this section, we start with the time evolution of laser energy absorption rate and give a three-stage theoretical model of laser-plasma interaction process. Fig.\ \ref{fig:figure2}\ shows the time evolution of laser energy absorption rate, in which the laser energy absorption rate refers to the ratio of the kinetic energy increment of electrons and ions to the increment of laser energy input during every 0.1 ps.

\begin{figure}[h]
  \includegraphics[scale=0.275]{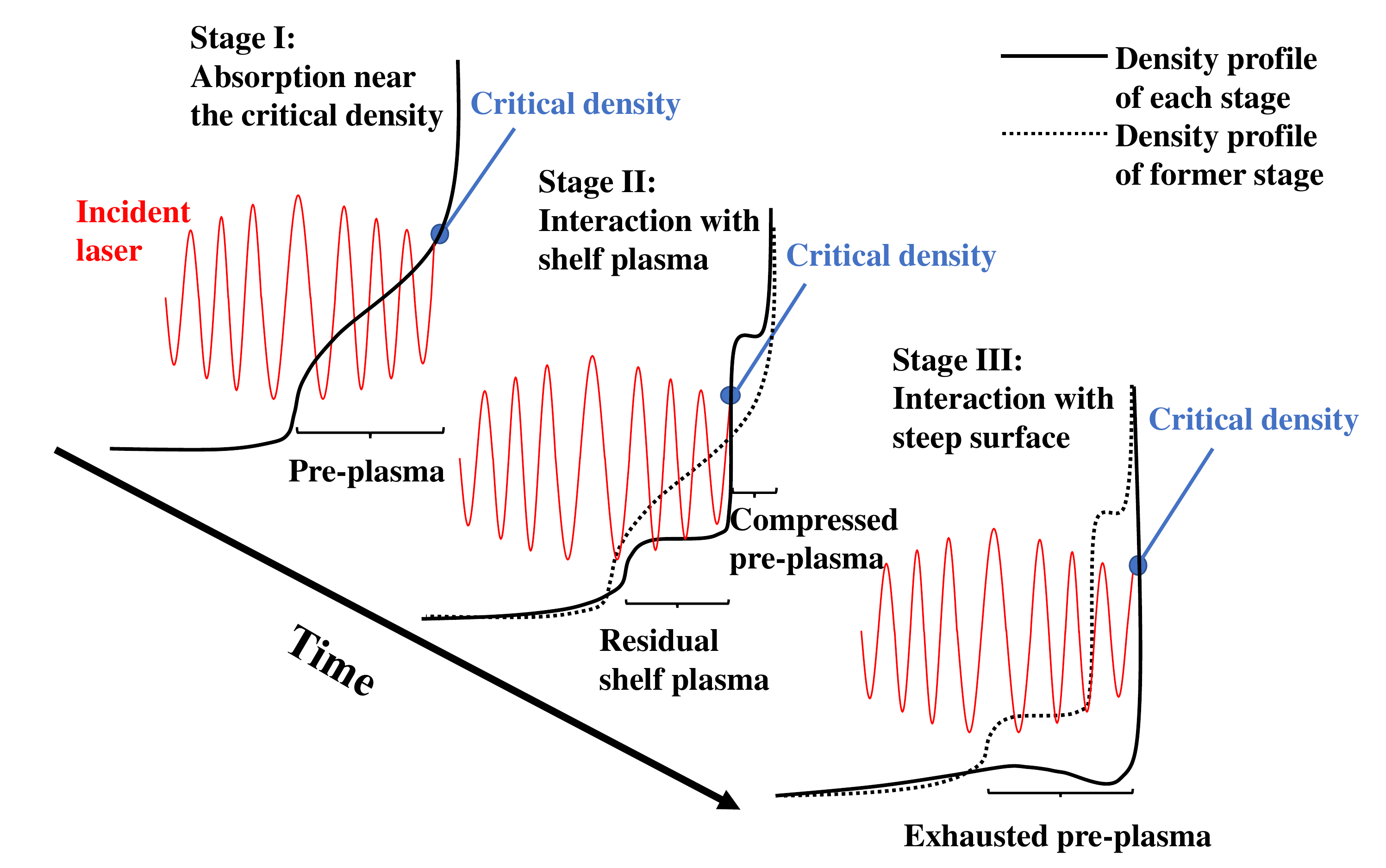}
  \caption{\label{fig:figure3} Schematic of the three-stage model of laser-plasma interaction. Stage I: Absorption near the critical density. Stage II: Interaction with shelf plasma. Stage III: Interaction with steep surface.}
\end{figure}

Note that at $t=0.2$ ps, the laser energy absorption rates in both cases are almost the same ($\sim17\%$), because the incident laser is interacting with the pre-plasma in front of the target which have the same density profile in different cases. During $t=0.3-0.5$ ps the laser energy absorption rates maintain at a relatively high value ($\sim27\%$). After $t=0.6$ ps, the energy absorption rates in both cases decrease significantly. Based on the above analysis, it can be inferred that the laser-target interaction process can be divided into three different stages \cite{r26,r27}, which is the reason for the change of laser energy absorption rate with time. A schematic of the three-stage model of laser-plasma interaction is shown in Fig. \ref{fig:figure3}.

\begin{figure}[h]
  \includegraphics[scale=0.25]{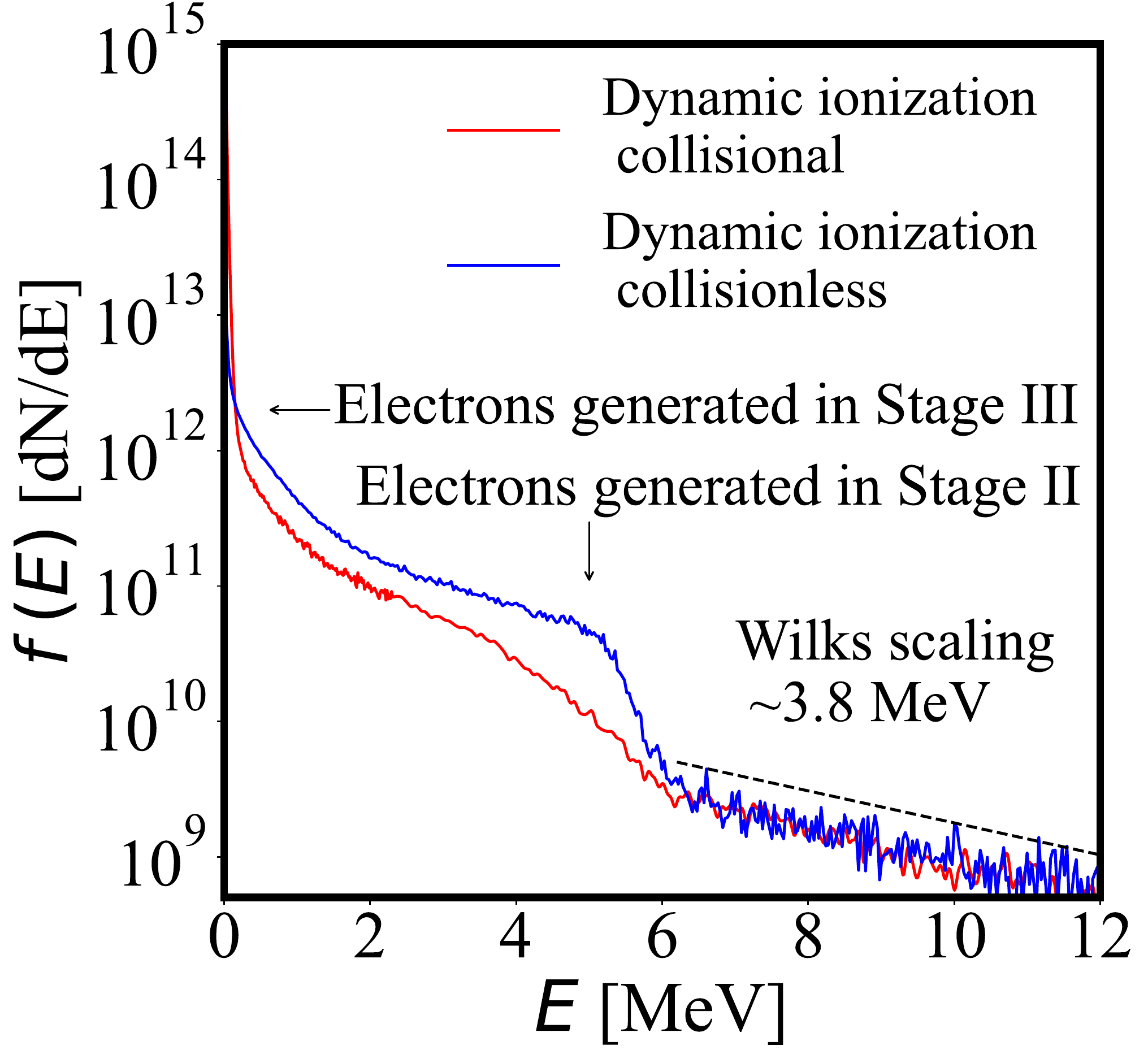}
  \caption{\label{fig:figure4} Time-integral energy spectrum of hot electrons passing through the front observation plane at $x=21\:\mu$m until $t=1.2$ ps.}
\end{figure}

\begin{figure}[h]
  \includegraphics[scale=0.375]{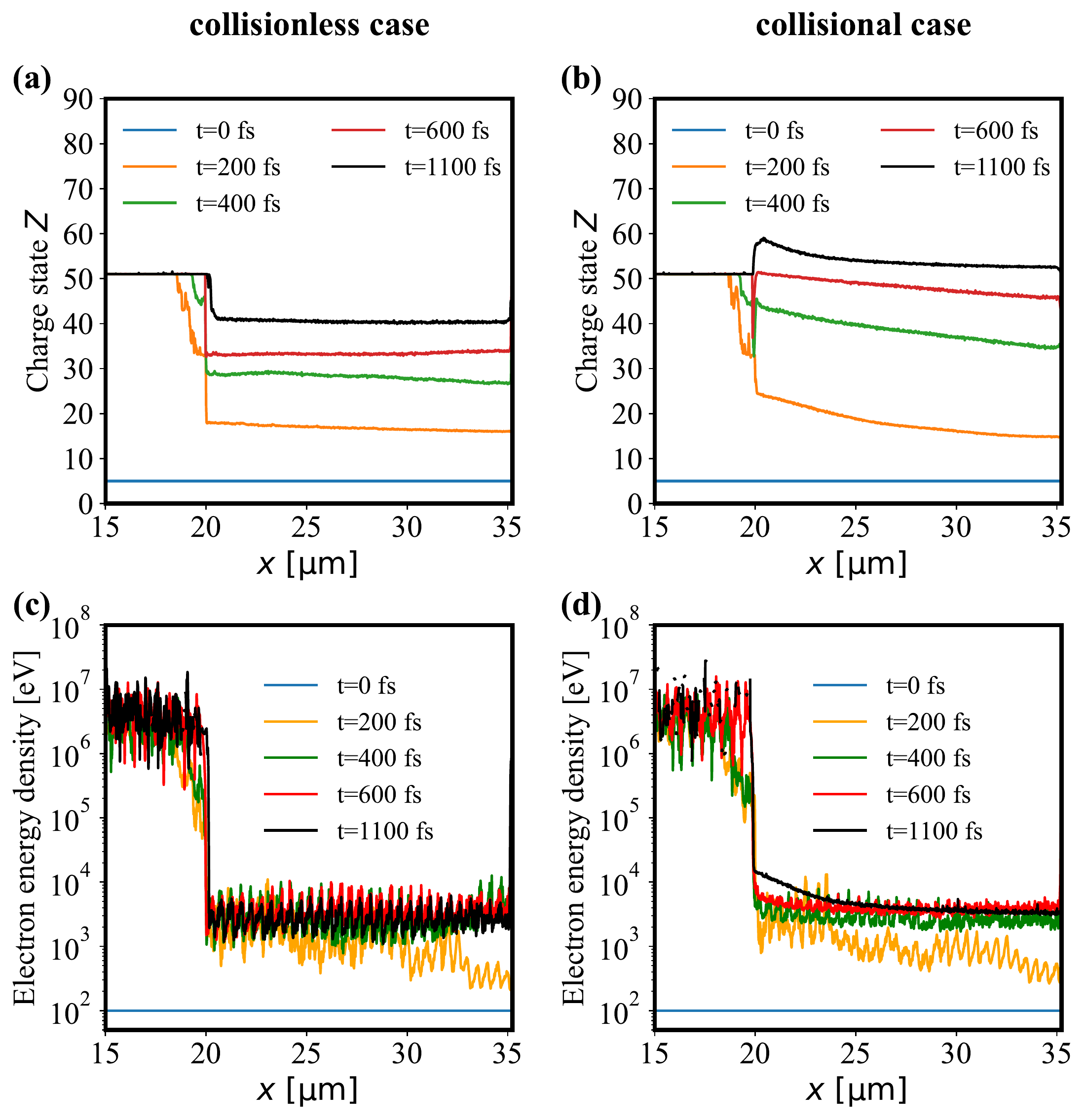}
  \caption{\label{fig:figure5} Time evolution of charge state of Au ions (a) (b) and energy density of hot electrons (c) (d) in the two cases considering dynamic ionization. (a) (c) Collisionless case, (b) (d) collisional case.}
\end{figure}

\begin{figure}[h]
  \includegraphics[scale=0.375]{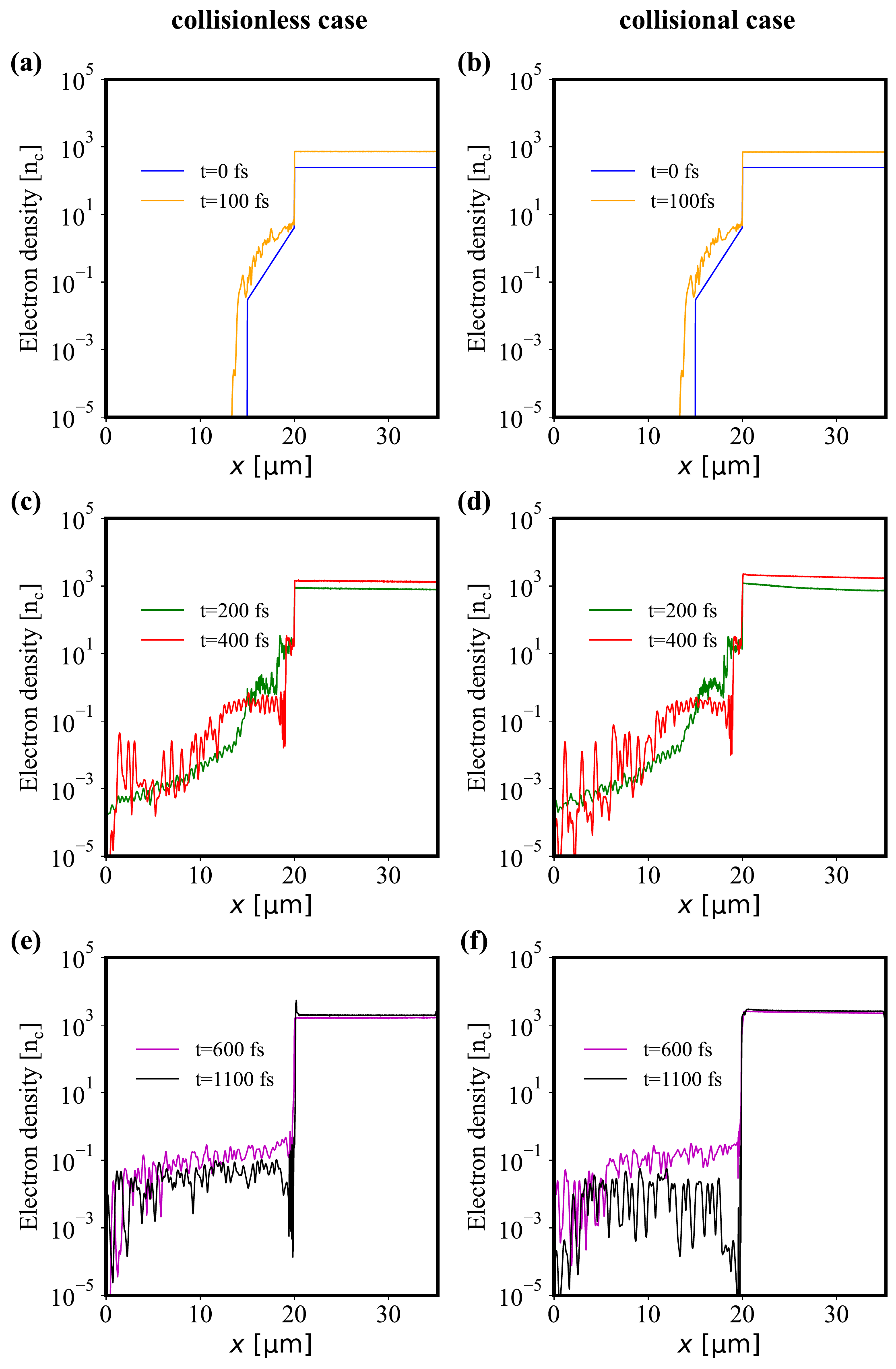}
  \caption{\label{fig:figure6} Time evolution of electron density in the two cases considering dynamic ionization. The first column represents collisionless case, and the second column represents collisional case. (a) (b) The first stage, (c) (d) the second stage, (e) (f) the third stage.}
\end{figure}

We label these three stages as (I) absorption near the critical density, (II) interaction with the shelf plasma, and (III) interaction with the steep surface, respectively. We will illustrate how these three stages affect laser energy absorption. In the first stage, the laser passes through the pre-plasma in front of the target and propagates to the relativistically modified critical density $n_c'=\gamma n_c\approx8.6n_c$, where $\gamma=\sqrt{1+I\lambda^2/1.37\times10^{18}}$ is the Lorentz factor. The dominant absorption mechanism in the first stage is stochastic heating and $J\times B$ heating in the pre-plasma. Fig.\ \ref{fig:figure4}\ shows the time-integral energy spectrum of hot electrons passing through the front observation plane at $x=21\:\mu$m for the two dynamic ionization cases until $t=1.2$ ps. As can be seen from Fig. \ref{fig:figure4}, the electrons in the energy spectrum are divided into different parts, corresponding to the three-stage model. The highest energy part corresponds to the first stage, and its slope (shown by the dotted line) is consistent with the temperature obtained by Wilks scaling \cite{r28} $\epsilon_T=(\gamma-1) m_e c^2 \sim 3.8$ MeV. Fig.\ \ref{fig:figure2}\ shows that the laser energy absorption rates of dynamic ionization cases with and without collision are almost the same before $t=0.4$ ps, which is consistent with the fact that the number of high-energy electrons ($E>6$ MeV) generated in the two cases are almost the same in Fig. \ref{fig:figure4}. This is due to the low collision frequency in the pre-plasma which has a low electron density and high temperature. The relationship between collision frequency $\nu$ and electron density $n_e$ and temperature $T_e$ is given by: $\nu \propto n_e T_e^{-3/2}$. From the temperature and electron density profile given in Fig. \ref{fig:figure5} and Fig. \ref{fig:figure6}, it can be estimated that the collision frequency in the pre-plasma is approximately seven orders of magnitude lower than that inside the target.

\begin{figure}[h]
  \includegraphics[scale=0.375]{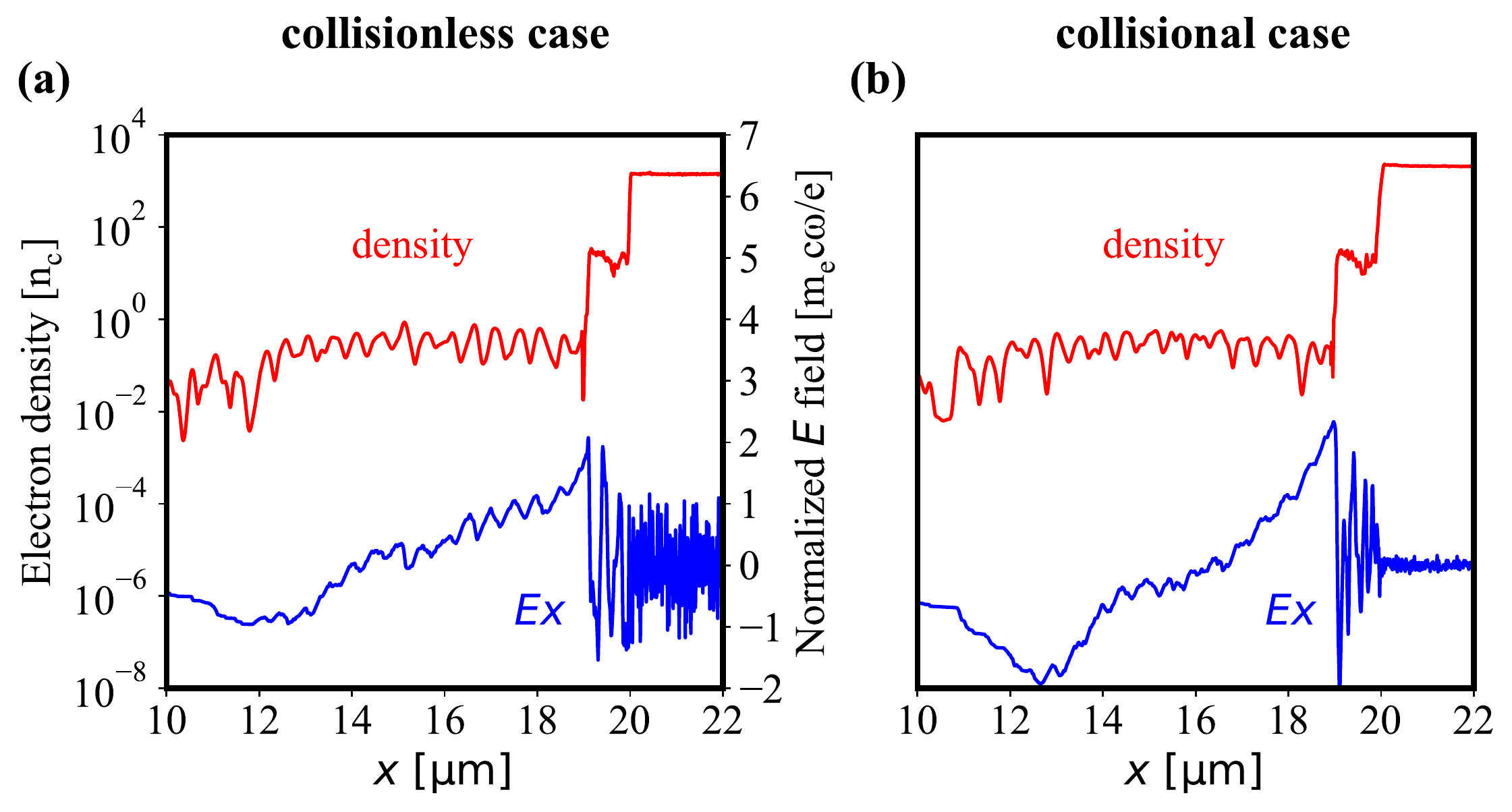}
  \caption{\label{fig:figure7} Electron density and the corresponding longitudinal electric field Ex of the two cases considering dynamic ionization at $t=400$ fs. (a) Collisionless case, (b) collisional case. The range of coordinates for the y-axis in (b) is the same as in (a), which we omit here for alignment with other figures.}
\end{figure}

\begin{figure}[h]
  \includegraphics[scale=0.31]{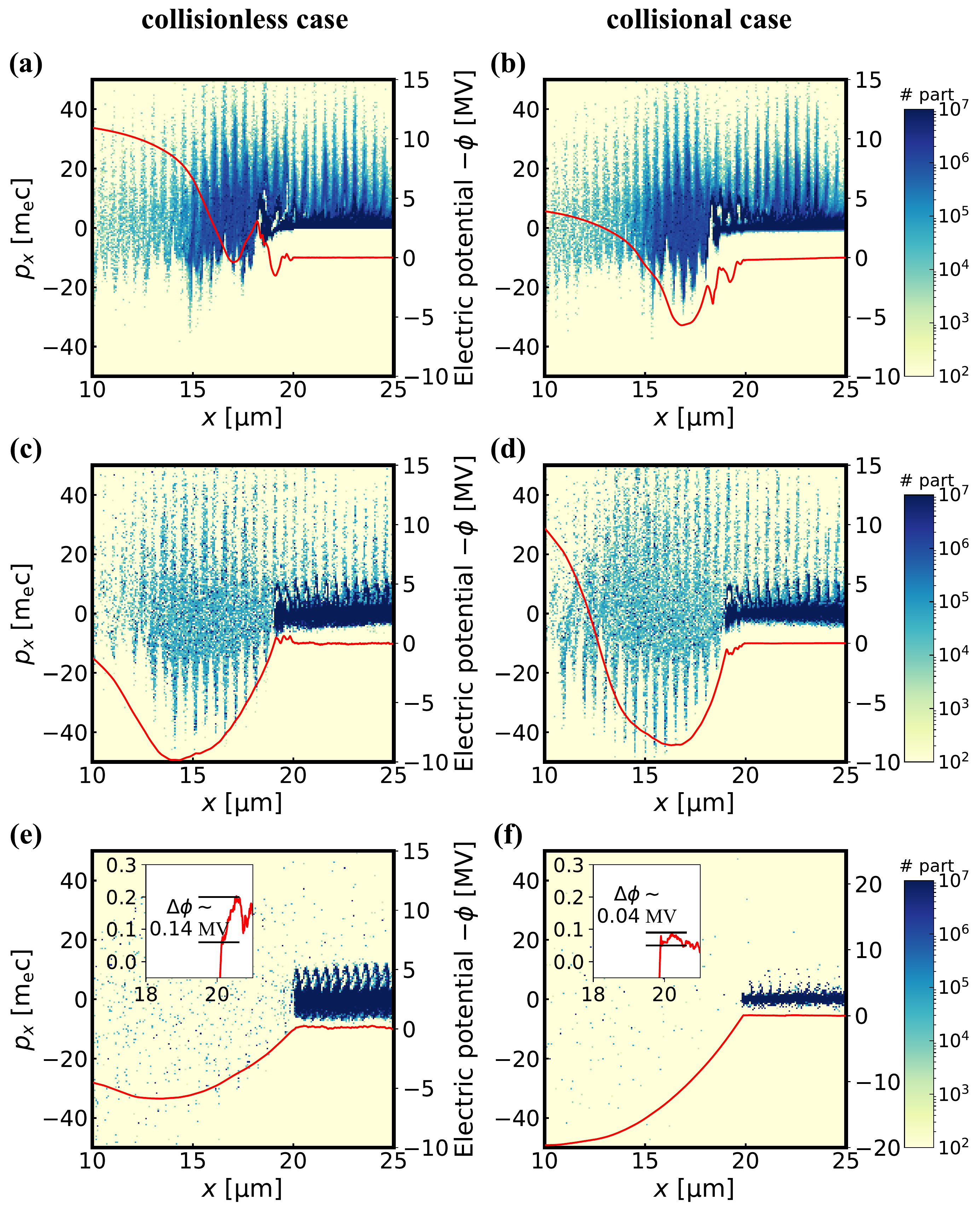}
  \caption{\label{fig:figure8} Electron phase plot and corresponding potential of the three stages. The first column (a) (c) (e) correspond to collisionless case and the second column (b) (d) (f) correspond to collisional case. (a) (b) $t=200$ fs, (c) (d) $t=400$ fs, (e) (f) $t=1100$ fs. The potential difference at the target surface is given in the small windows of (e) (f).}
\end{figure}

Then, under the action of laser ponderomotive force, pre-plasma is gradually compressed into the target and the ions are ionized to produce more free electrons, symbolizing the second stage of laser-plasma interaction. Fig.\ \ref{fig:figure5}\ shows the time evolution of Au ions charge state in the two cases considering dynamic ionization and Fig.\ \ref{fig:figure6}\ shows the time evolution of electron density in the two cases. Due to field ionization, the charge state of the ions in front of the target quickly increases to $Z=51$ and remains unchanged. For the collisionless case, the charge state of Au ions in the target is relatively uniform. However, for the collisional case, the energy deposition of fast electrons causes the local charge state of several $\mu$m depths in the target to be higher. This also results in an overall higher charge state of the target than in the collisionless case. Fig.\ \ref{fig:figure7}\ shows the electron density and the corresponding longitudinal electric field Ex of the two dynamic ionization cases at $t=400$ fs. Electrons are gathered in front of the target by strong positive-negative field, and are then loaded into the vacuum by another electric field on the left. Fig.\ \ref{fig:figure8}\ shows the electron longitudinal phase plot and corresponding electrostatic potential for the three stages. Note that in the figures, $-\phi$ is plotted instead of $\phi$ in order to compare with the energy of electrons. As shown in Fig. \ref{fig:figure8}(c) and (d), an electrostatic potential well is formed before the target in the second stage of laser-plasma interaction. It is known that the electron oscillating coherently with the electric field of a plane wave obtains zero energy in one period. However, due to the electrostatic potential in front of the target, the phase coherence is broken when electrons moving in the electrostatic potential well \cite{r27}, so that electrons can be accelerated to an energy much higher than the maximum electrostatic potential by the incident laser and reflected laser. The spectrum of the electrons generated in the second stages form a shoulder-like area in Fig. \ref{fig:figure4}. Ionization not only provides an additional energy absorption mechanism, but also generates more electrons for acceleration, leading to an increase in the laser energy absorption rate in this stage. During pre-plasma compression, on the one hand, collision increases the charge state and charge to mass ratio ($Z/A$) of Au ions, making the ions and pre-plasma in front of the target easier to be swept away; on the other hand, collision provides a hot pressure $P=2n_eT_e/3$ to counteract the laser ponderomotive force. The pre-plasma compression processes with and without collision are almost identical due to the two opposite effects of collision. It can be seen from Fig.\ \ref{fig:figure6}\ that in both collisionless case and collisional case, the pre-plasma is swept away under the action of the laser ponderomotive force at $t=600$ fs, leading to the third stage of laser-plasma interaction.

The laser directly interacts with the target surface and is reflected at the critical density $\gamma n_c$ in the third stage, whose energy absorption rate and the electron energy are the lowest of the three stages. Since the pre-plasma in front of the target is almost completely swept away, electron acceleration in the electrostatic potential well is no longer important in the third stage. As shown in Fig. \ref{fig:figure8}(e) and (f), only a small number of electrons escape from the front surface of the target. Ignoring the small fraction of electrons escaping from the target surface, there are equilibrium conditions for pressure, electron flux, and energy at the target surface. First, the pressure equilibrium condition for electrons at the laser critical interface can be written as:
\begin{equation}
    \frac{(1+\eta)I_{in}}{c} = n_ee\Delta\phi+\frac{2}{3}n_eT_e
\end{equation}

In which $\eta, I_{in}$ are the laser reflectivity and laser intensity, $n_e, T_e$ are the density and temperature (in the energy unit) of electrons, $\Delta\phi$ is the potential difference of the charge separation field caused by laser light pressure. For collisionless cases, the last term $2n_eT_e/3$ does not exist. Therefore potential difference $\Delta\phi$ in collisionless cases will be larger than that in collisional cases. As is shown in Fig. \ref{fig:figure8}(e) and (f), the potential difference in collisionless case is $0.1$ MV higher than that in collisional case, which is consistent with the interface temperature of about $0.1-1$ MeV in collisional case shown in Fig. \ref{fig:figure5}(d). The fast electron energy can be written as:
\begin{equation}
    \epsilon_e=(\sqrt{1+\frac{{p_x}^2}{m_e^2c^2}}-1)m_ec^2 = \frac{(1+\eta)I_{in}}{n_ec}
\end{equation}

And the equilibrium conditions for electron flux and energy flux are given by:
\begin{equation}\label{eq3}
    n_rv_r=n_fv_f
\end{equation}

\begin{equation}
    I_{in}-I_{re}=\chi I_{in}=\epsilon_ev_fn_f
\end{equation}

In which $n_r, v_r$ are the density and velocity of electrons which returns to the target surface from inside the target, $n_f, v_f$ are the density and velocity of fast electrons that gain energy at the surface and are re-injected into the target. $I_{in}$ and $I_{re}$ are the intensity of incident laser and reflected laser while $\chi$ represents the laser energy absorption rate. From equation(2)-(4), the absorption rate is obtained:

\begin{equation}
    \chi\approx\frac{2}{a}\frac{n_f}{n_c}(1-\frac{1}{\gamma_f})\sqrt{\frac{n_c}{n_e}}
\end{equation}

Normalized laser amplitude $a=\sqrt{I\lambda^2/1.37\times10^{18}}$ and factor $\gamma_f\sim1+a^2(n_c/n_e)$. In the collisional cases, the velocity of the return electrons $v_r$ are reduced, resulting in a decrease in the density of the generated fast electrons $n_f$. Meanwhile, due to the decrease in electric potential difference $\Delta\phi$, the energy of the fast electrons $\epsilon_e$ also decreases in collisional cases. The above are the reasons why the laser energy absorption rate decreases due to collision in the third stage. These conclusions can be clearly seen in Fig \ref{fig:figure8}\ (e) and (f).

\maketitle
\section{Effect of target charge state on fast electron generation}
\begin{figure}[h]
  \includegraphics[scale=0.375]{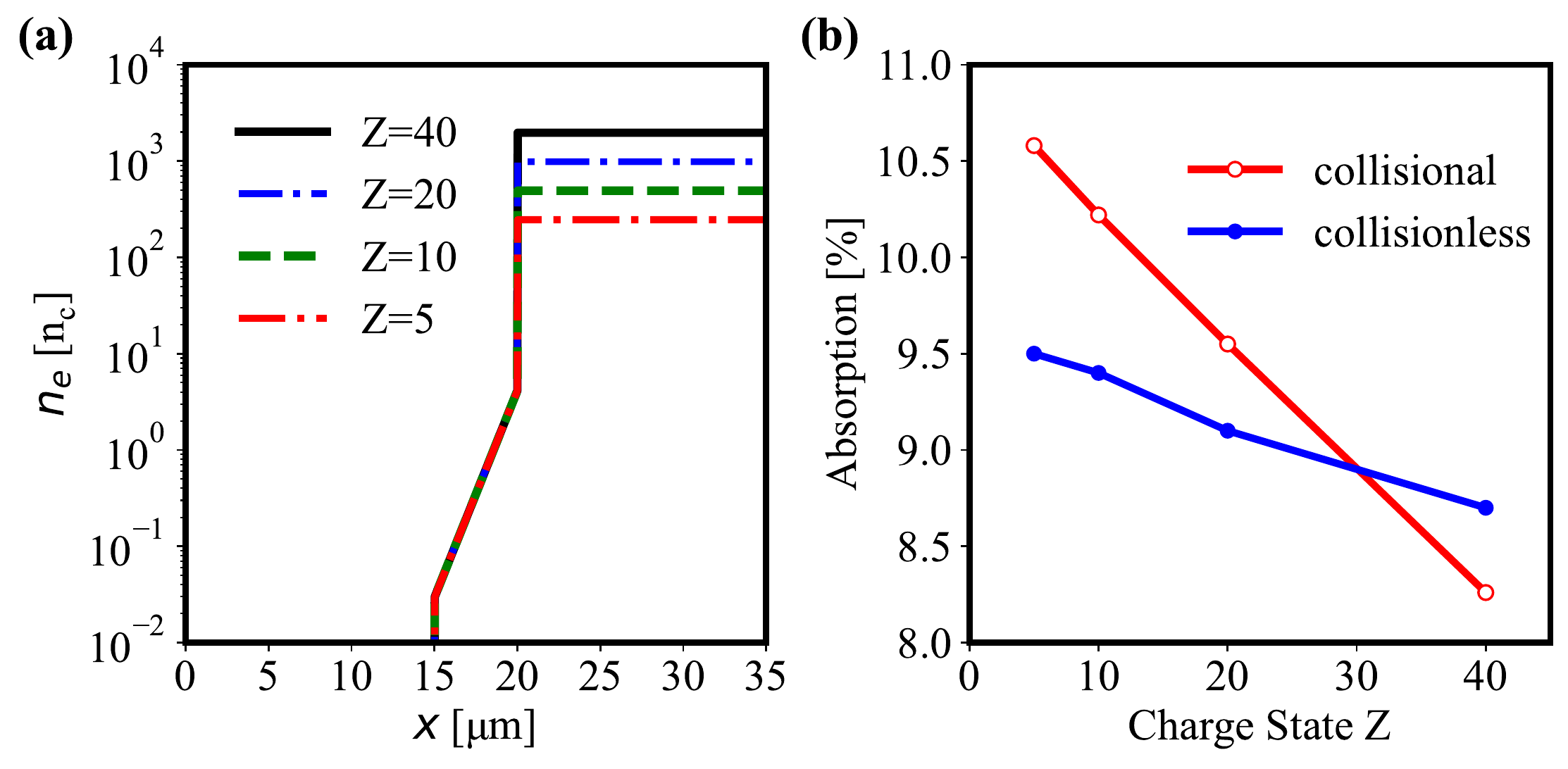}
  \caption{\label{fig:figure9} (a) Initial electron density profile of cases with different charge state of Au ions. (b) The relationship between laser energy absorption rate and charge state Z. The red line and the blue line represent collisional and collisionless cases with different charge state, respectively.}
\end{figure}

In this section we discuss how the charge state of Au ions affects laser energy absorption and fast electron generation. To control the variables, we performed simulations with different fixed charge states whose electron density profiles are shown in Fig. \ref{fig:figure9}(a). As can be seen from Fig. \ref{fig:figure9}(b), in both collisional and collisionless cases, laser energy absorption rate decreases with the increasing of charge state Z of Au ions. Here absorption rate refers to the ratio of the total kinetic energy of fast electrons and ions generated in the simulation to the total incident laser energy.

\begin{figure}[h]
  \includegraphics[scale=0.375]{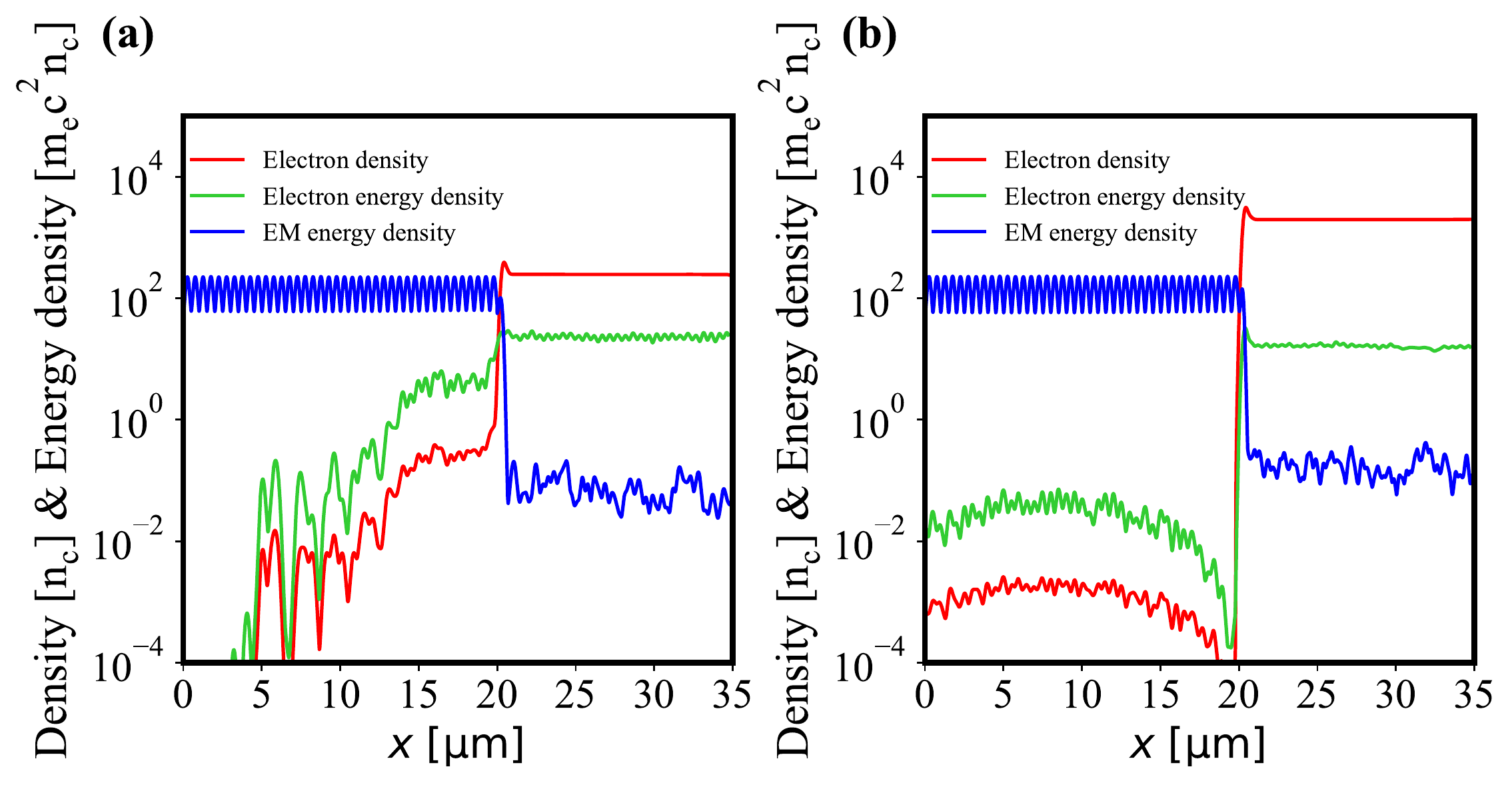}
  \caption{\label{fig:figure10} Spatial profile of (red) electron density, (green) electron energy density, and (blue) laser energy density for collisionless cases (a) $Z=5$ and (b) $Z=40$ at $t=1$ ps.}
\end{figure}

\begin{figure}[h]
  \includegraphics[scale=0.31]{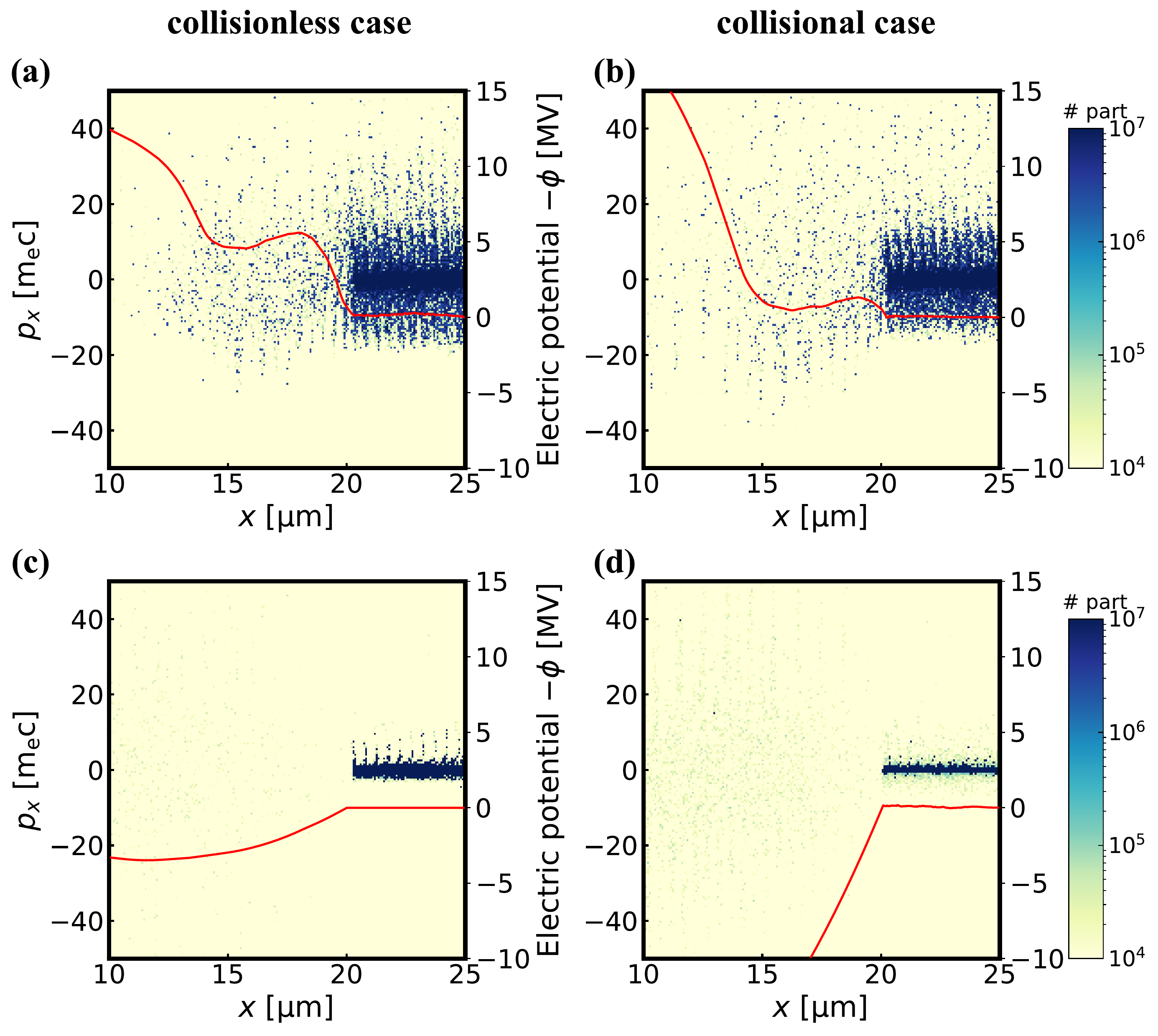}
  \caption{\label{fig:figure11} Electron phase plot and corresponding potential at $t=1.1$ ps for (a) $Z=5$ collisionless case, (b) $Z=5$ collisional case, (c) $Z=40$ collisionless case, (d) $Z=40$ collisional case.}
\end{figure}

\begin{figure}[h]
	\includegraphics[scale=0.375]{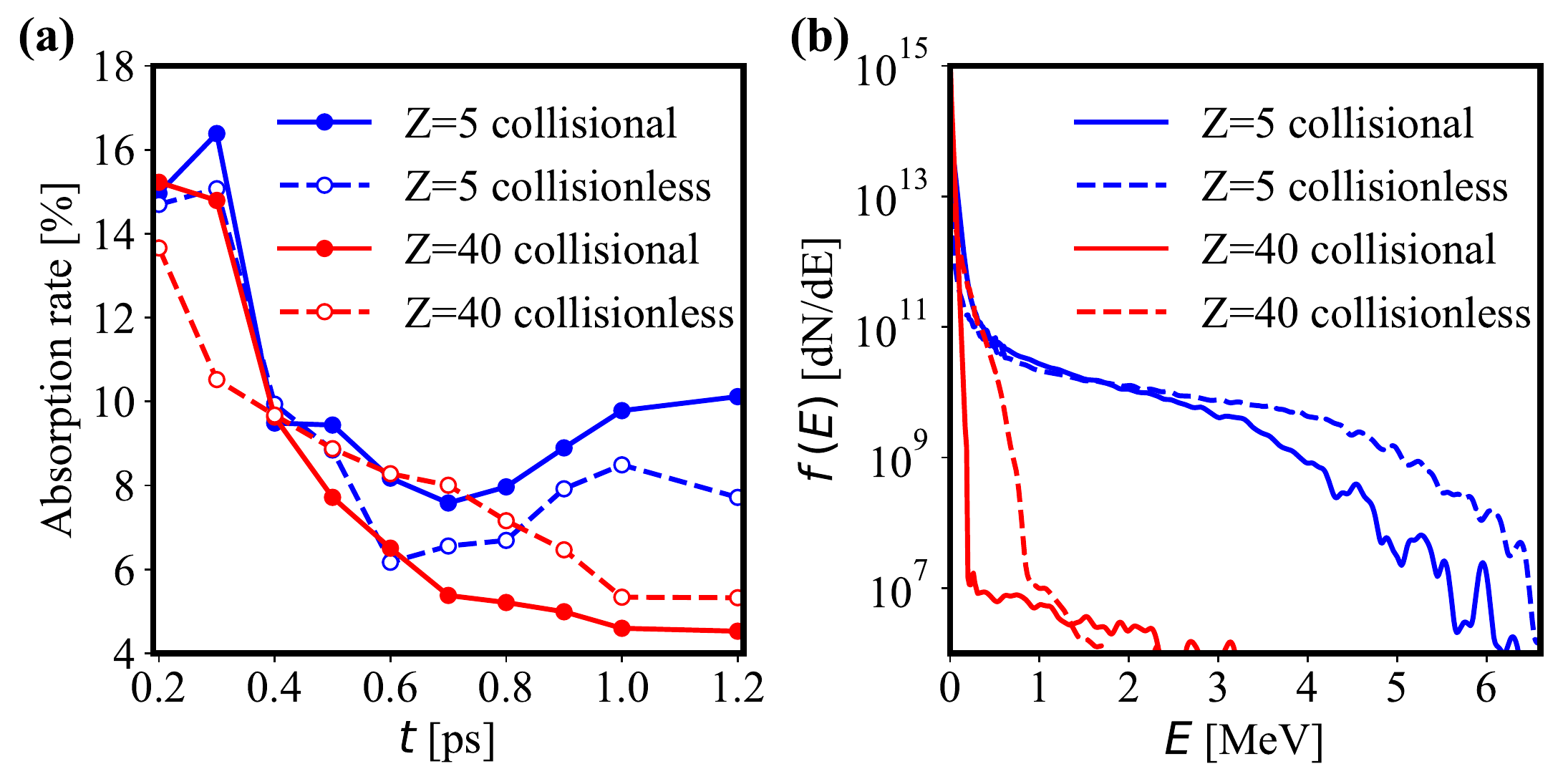}
	\caption{\label{fig:figure12} (a) Time evolution of laser energy absorption rate under different conditions with fixed charge state of Au ions calculated every $0.1$ ps. (b) Fast electron spectrum at $t=1.1$ ps for (solid line) collisional cases and (dashed line) collisionless cases with (blue) $Z=5$ and (red) $Z=40$.}
\end{figure}

In order to better understand the reason for the decrease in laser energy absorption rate and verify the three-stage model in the previous section, Fig. \ref{fig:figure10}(a) and (b) show the spatial profile of electron density, electron energy density, and the laser energy density corresponding to the charge state $Z=5$ and $Z=40$ at $t=1$ ps, respectively. When the laser is incident, it will compress the pre-plasma into the target as is discussed in the previous section. During the compression of pre-plasma, electrons in the pre-plasma are pushed into the Au target by laser ponderomotive force, forming an charge separation field between these electrons and remaining Au ions to push the ions into the Au target. For cases with higher charge state Z, the ions have a higher charge to mass ratio Z/A and are easier to move under the electric field. Therefore, in the high-Z cases, the pre-plasma is quickly swept away, allowing the laser to interact with the solid target directly. As is shown in Fig. \ref{fig:figure10}, for $Z=5$ case there is still pre-plasma with an electron density $\sim1n_c$ in front of the target which means that the laser-plasma interaction still remain in the second stage for $Z=5$ case, while for $Z=40$ case the pre-plasma is almost swept away completely which means that the laser-plasma interaction has entered the third stage. Fig.\ \ref{fig:figure11}\ shows the electron longitudinal phase plot and corresponding electrostatic potential at $t=1.1$ ps for $Z=5$ cases (a) (c) and $Z=40$ cases (b) (d). As expected, from the phase plot and potential, the two cases of $Z=5$ remain in the second stage, while the two $Z=40$ cases has entered the third stage. Fig.\ \ref{fig:figure12}\ shows the dynamic laser energy absorption rate calculated every 0.1 ps. It can be seen that the dynamic laser energy absorption rate of $Z=40$ case decreases significantly after $t=0.4$ ps while the dynamic laser energy absorption rate of $Z=5$ case remains relatively high. Meanwhile less fast electrons are generated in $Z=40$ case than in $Z=5$ case as shown in Fig. \ref{fig:figure12}(b). These results indicate that a fixed low charge state of the high-Z target which is usually assumed in the simulations is not valid. This is due to the fact that charge state will affect the evolution into the third stage and that the late stage absorption rate of the fixed low charge state will be very different from the fixed high charge state case and the dynamic ionization case.

\maketitle
\section{Collisional effects on laser-target interaction}
In this section, collisional effects will be discussed, revealing how the presence of collision affects laser-target energy coupling in different stages. The `collision' here refers to the binary Coulomb collision between charged particles, which is calculated in a natural manner based on the Monte Carlo method in our PIC code. 

As is shown in Fig. \ref{fig:figure9}(b), absorption rates in collisional cases are higher than those in collisionless cases except for the cases of $Z=40$. Our results show that collision has both positive and negative effects on the laser energy absorption rate. The positive effect of collision is to prevent the laser from compressing the pre-plasma and make it easier for the electrons to diffuse into the vacuum in front of the target. Due to the presence of collision, there will be a thermal pressure $2n_eT_e/3$ generated by the thermal motion of the electrons and the Coulomb force between the electrons. The coefficient $2/3$ is due to $T_e=E_{\mathrm{hot}}=3k_BT/2$, in which $E_{\mathrm{hot}}$ is electron average kinetic energy, $k_B$ and $T$ are Boltzmann constant and temperature of electrons, respectively. As is shown in Fig. \ref{fig:figure11}(a) and (b), for the second stage of laser-plasma interaction ($Z=5$), it is easier for electrons in the collisional case to escape into the electric potential well and be accelerated, resulting in more fast electron generation and higher energy absorption rate. Therefore, in the second stage of laser-plasma interaction, collision is beneficial for laser-target energy coupling.

The negative effect of collision is causing electrons in the return current scattered by ions. As is shown in Fig. \ref{fig:figure11}(c) and (d), in the collisionless case of $Z=40$, both the forward and return electron beams are more energetic. This is consistent with the result in Fig. \ref{fig:figure9}(b) that shows a higher absorption rate in collisionless case than that in collisional case of $Z=40$. In the collisionless case, electrons can travel back to the front surface of Au target through the return current and re-reach the LPI region. While in the collisional case, the electrons in the return current are scattered by the ions, which lowers the velocity of return electrons $v_r$ in Eq. \ref{eq3}. So collision is harmful for laser-target energy coupling in the third stage of laser-plasma interaction.

In summary, there are abundant electrons in the shelf plasma to be accelerated in the second stage, in which collision helps electrons diffuse into the electric potential well and be accelerated by the laser; however, there is almost no pre-plasma left in front of the target in the third stage so the supply of electrons mainly depends on the return current within the target which will be severely suppressed by collision. Therefore, though collision is beneficial for laser energy absorption in the second stage, it will be harmful in the third stage. And it can be inferred from Fig. \ref{fig:figure9}(b) whether the laser-plasma interaction will enter the third stage is mainly determined by the charge state of Au ions.

\begin{figure}[h]
	\includegraphics[scale=0.25]{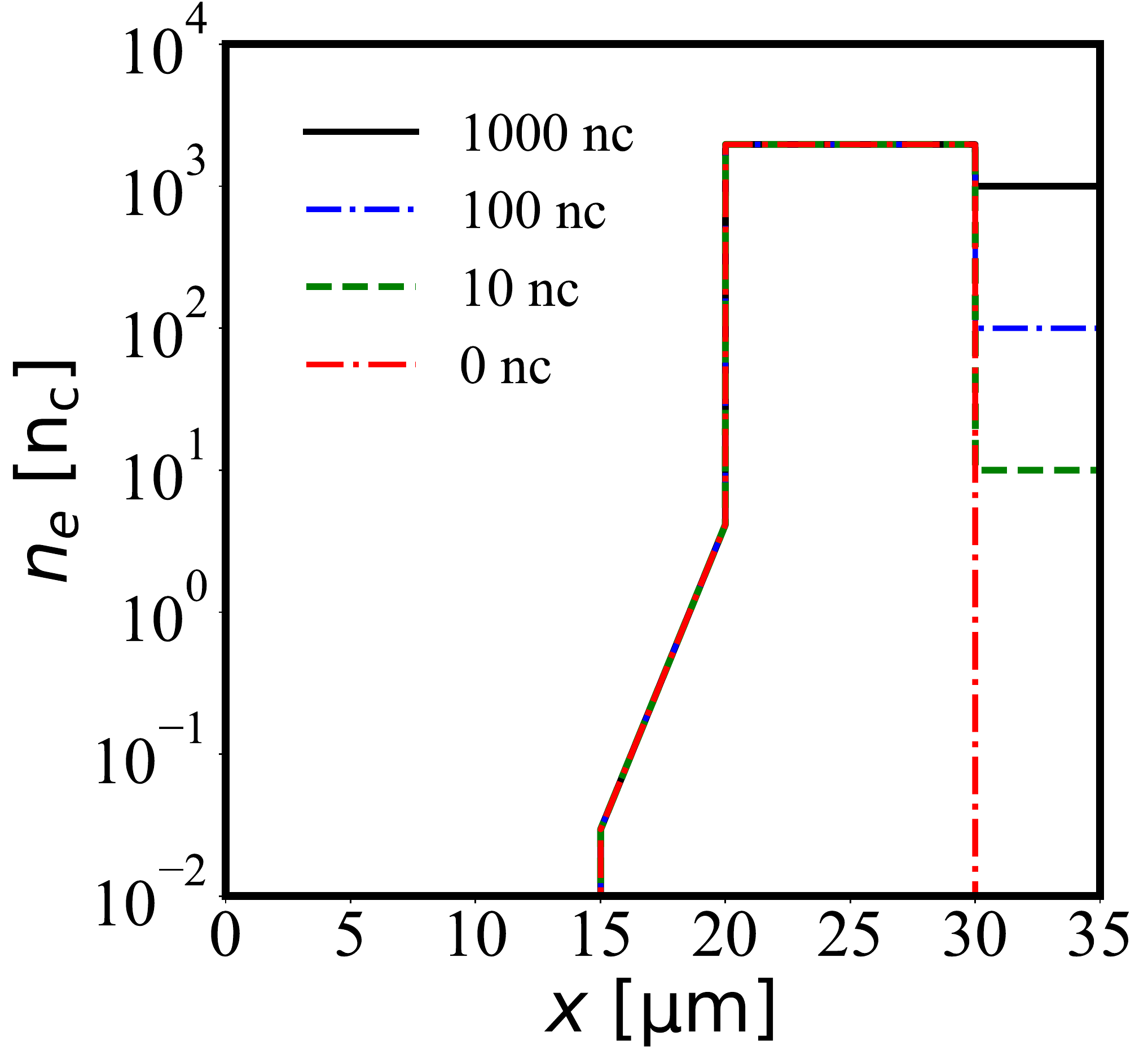}
	\caption{\label{fig:figure13} Initial electron density profile of cases with different blow-off plasma densities.}
\end{figure}

\begin{figure}[h]
	\includegraphics[scale=0.375]{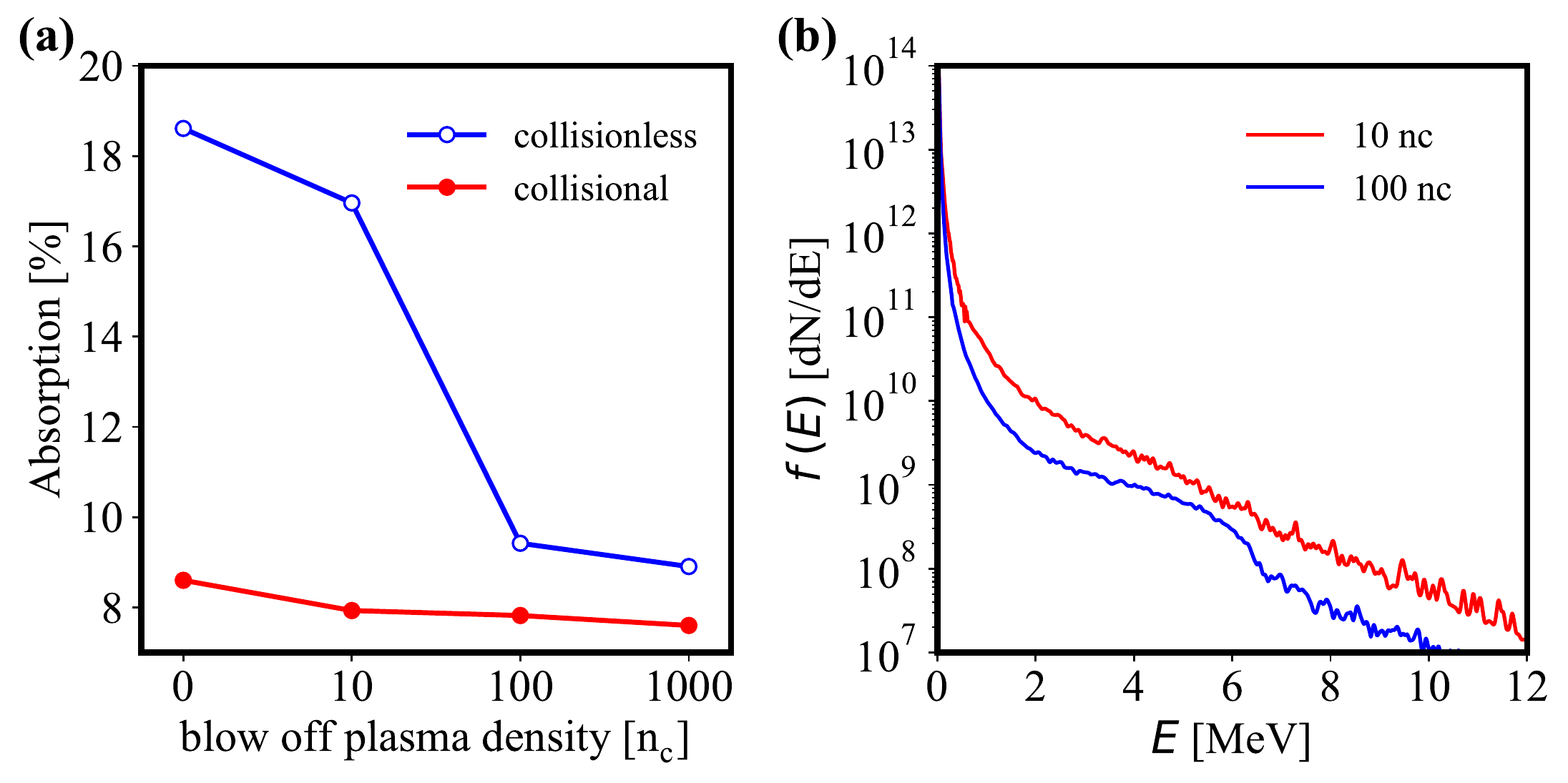}
	\caption{\label{fig:figure14} (a) Laser energy absorption rate in (blue) collisionless cases and (red) collisional cases with different blow-off plasma densities. (b) Fast electron spectrum of two collisionless cases (red) 10$n_c$ and (blue) 100$n_c$ at $t=1$ ps.}
\end{figure}

\begin{figure}[h]
	\includegraphics[scale=0.375]{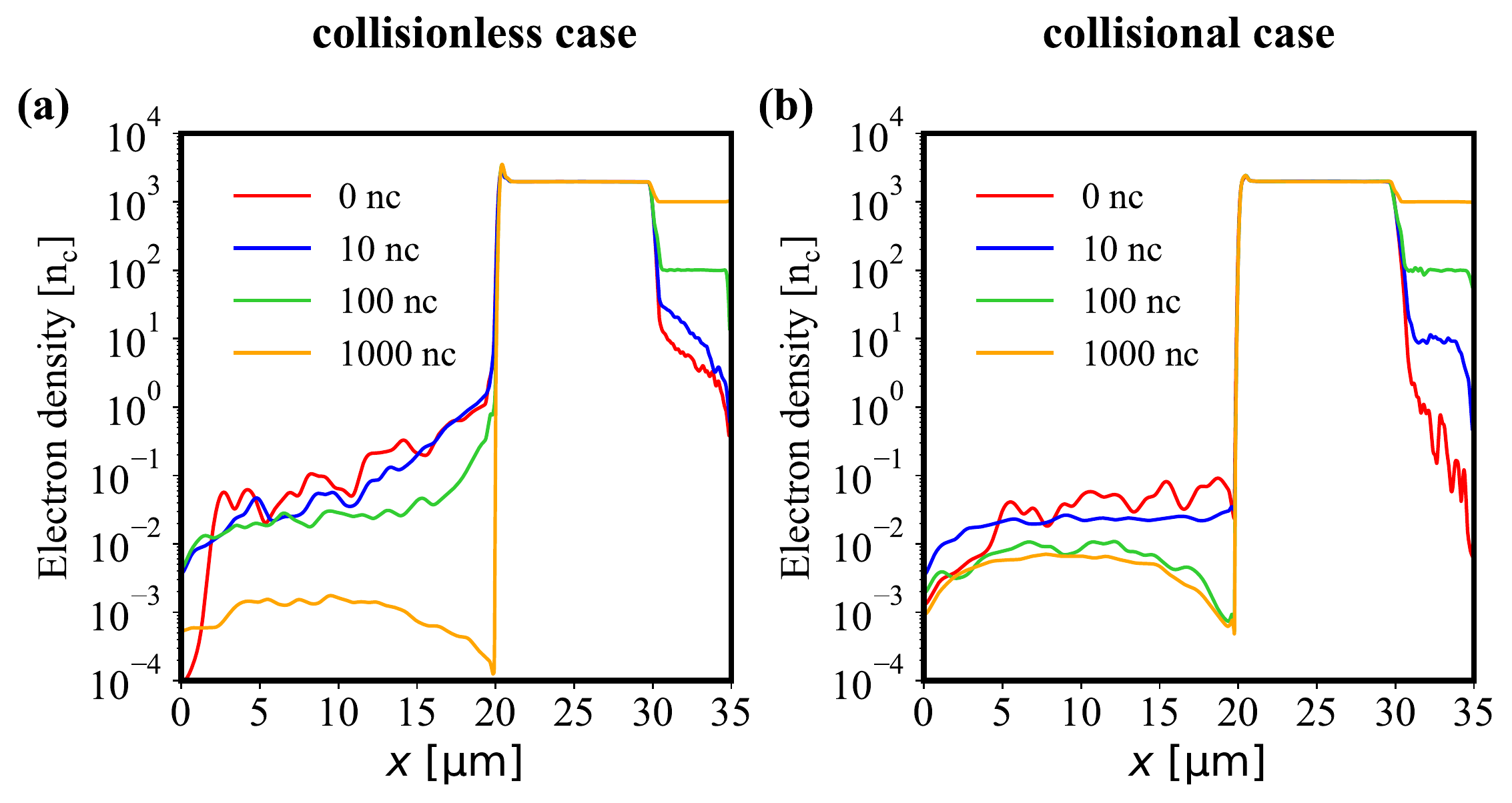}
	\caption{\label{fig:figure15} Electron density profile at $t=1$ ps for (a) collisionless and (b) collisional cases from 0$n_c$ to 1000$n_c$.}
\end{figure}

\begin{figure}[h]
	\includegraphics[scale=0.375]{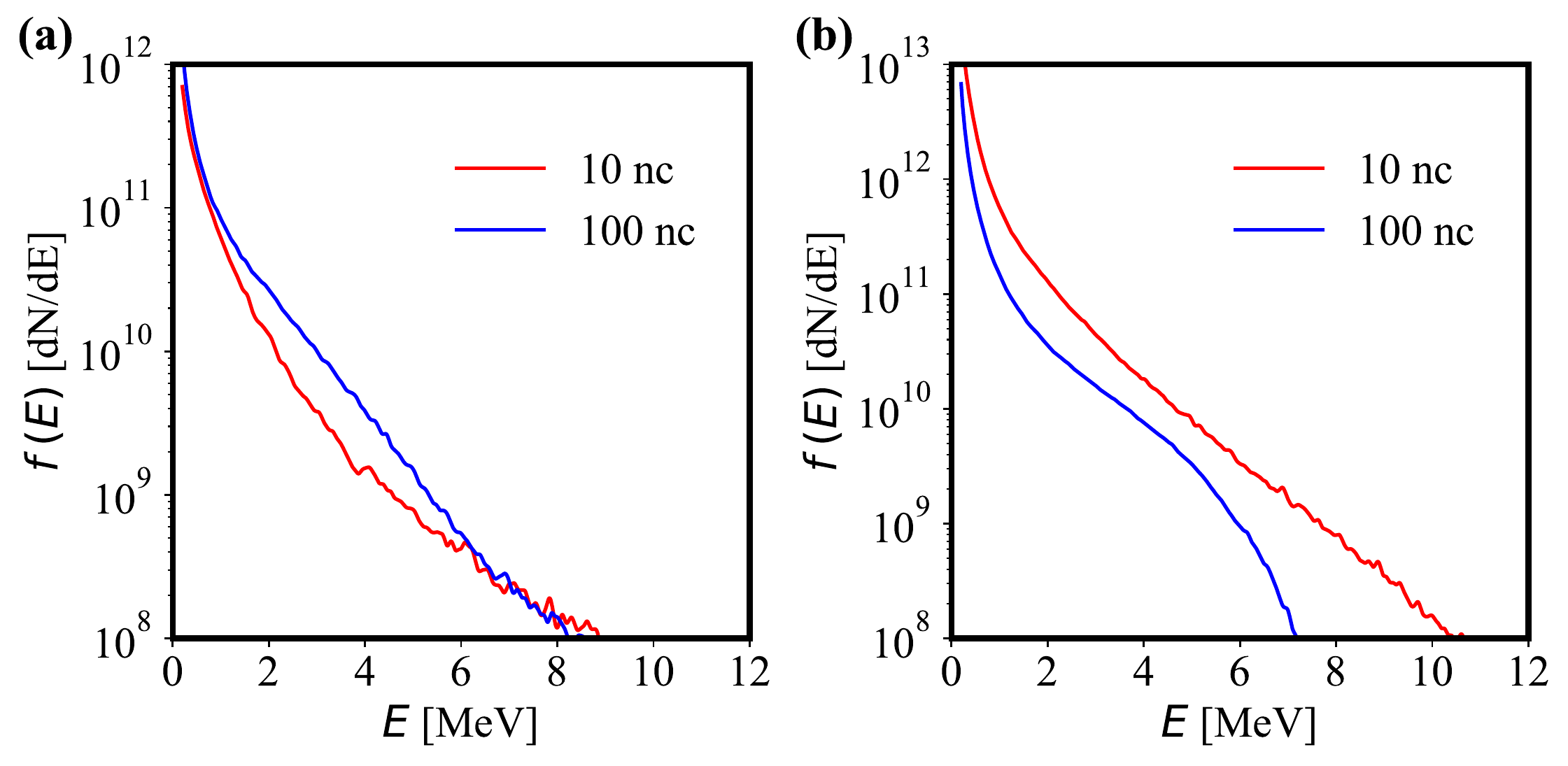}
	\caption{\label{fig:figure16} (a) Spectrum of electrons passing forwards through the rear observation plane at $x=34\:\mu$m; (b) Spectrum of electrons passing backwards through the front observation plane at $x=21\:\mu$m.}
\end{figure}

\begin{figure}[h]
	\includegraphics[scale=0.325]{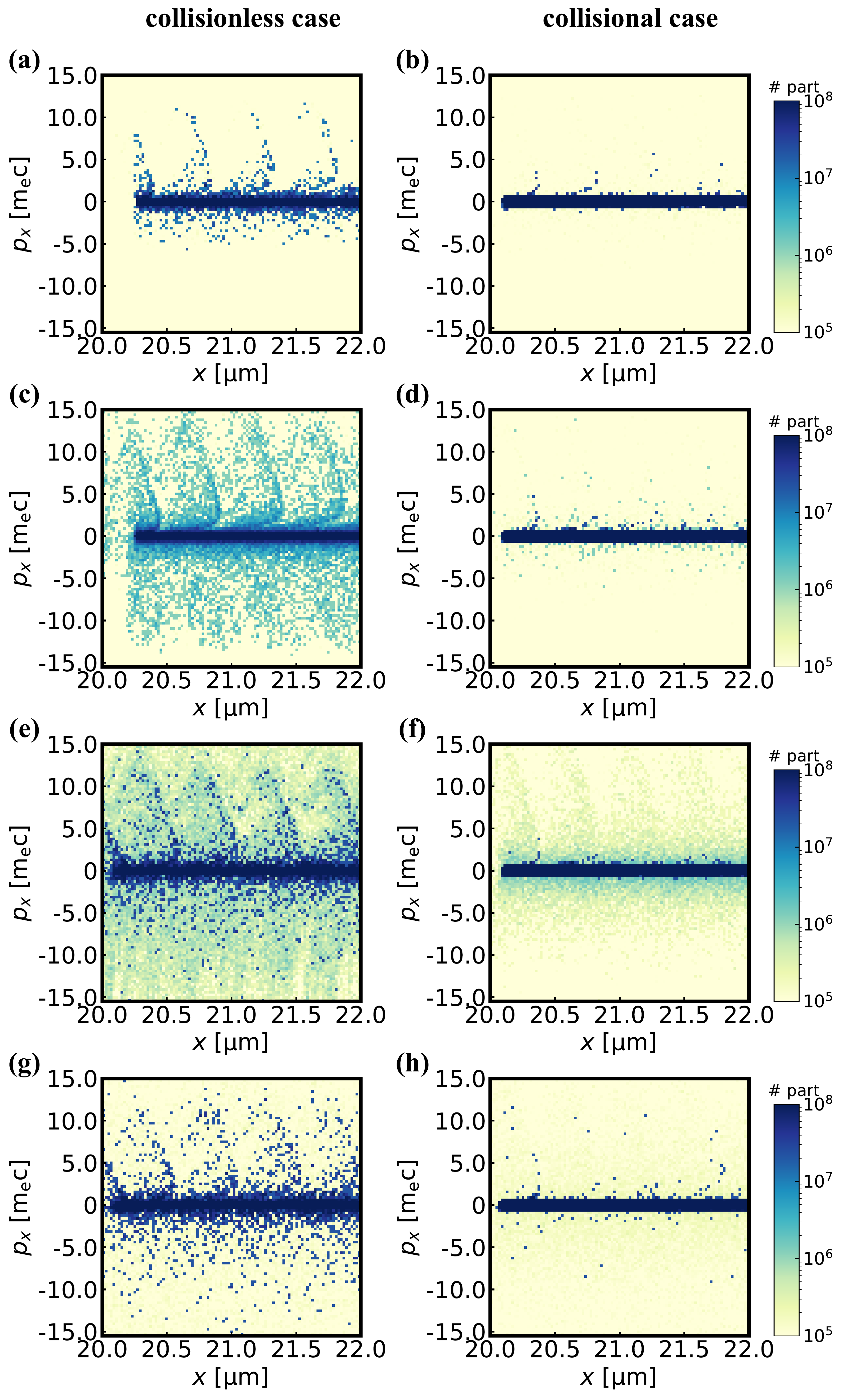}
	\caption{\label{fig:figure17} Electron longitudinal phase plot at $t=1$ ps for collisionless cases (the first column) and collisional cases (the second column) with different blow-off plasma densities. (a) (b) $1000n_c$; (c) (d) $100n_c$; (e) (f) $10n_c$; (g) (h) $0n_c$.}
\end{figure}

\begin{figure}[h]
	\includegraphics[scale=0.325]{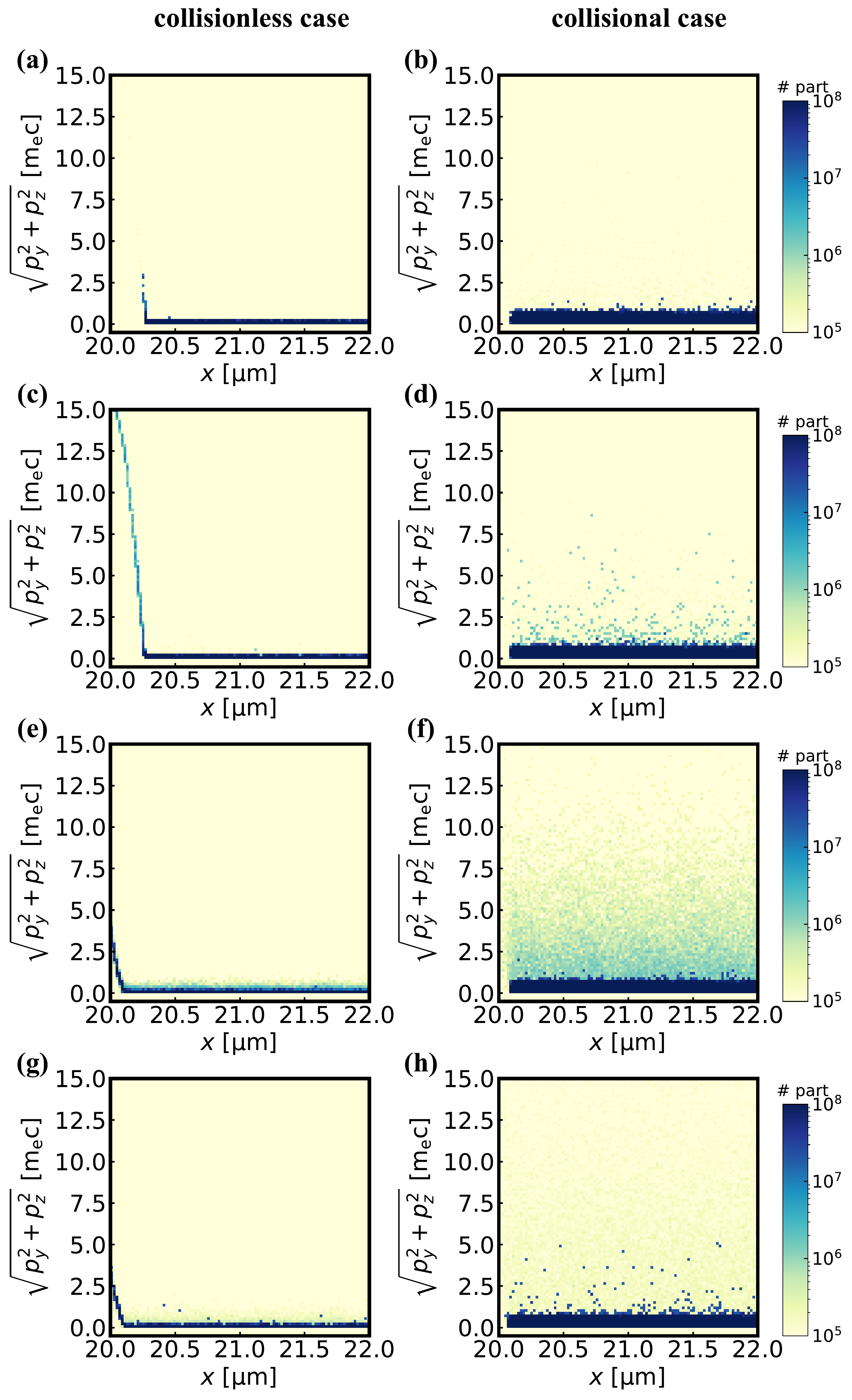}
	\caption{\label{fig:figure18} Electron transverse phase plot at $t=1$ ps for collisionless cases (the first column) and collisional cases (the second column) with different blow-off plasma densities. (a) (b) $1000n_c$; (c) (d) $100n_c$; (e) (f) $10n_c$; (g) (h) $0n_c$.}
\end{figure}

\maketitle
\section{collisional effects on electron recirculation in the presence of blow-off plasma}
Notice that the Au cone is typically inserted into plasma, which may influence the generation of fast electrons. Therefore cases in which there are several micro meters of blow-off plasma behind the target should also be discussed. In order to study the effect of blow-off plasma on laser energy absorption and electron generation, the 5 $\mu$m thickness end of the target is replaced by blow-off plasma of different densities in the simulation. The initial density of the target is shown in Fig. \ref{fig:figure13}. Here the target charge state is fixed as $Z=40$ and only the blow-off plasma density differs between different cases. Hereinafter each case will be referred to as its blow-off plasma density. To verify that the gap of 5 $\mu$m behind the target is long enough, we added simulations in which the blow-off plasma density is still 0 $nc$ but the gap behind the target is increased to 15 $\mu$m. It is found that the laser energy absorption rates of 5 $\mu$m gap and 15 $\mu$m gap are almost the same. This is due to the fact that the 5 $\mu$m gap behind the target has already prevented most of electrons from leaving the right boundary. Therefore, it can be considered that the gap of 5 $\mu$m behind the target is long enough.

It is believed that the blow-off plasma has the effect of avoiding fast electron recirculation \cite{r15}. Fig. \ref{fig:figure14}(a) shows the laser energy absorption rate over the whole simulation duration of both collisionless and collisional cases with different blow-off plasma densities. With the increasing of blow-off plasma density, the absorption rates of collisionless cases decrease significantly. The energy spectrum in Fig. \ref{fig:figure14}(b) also shows that less fast electrons are generated in the collisionless $100n_c$ case than in the $10n_c$ case.

In Fig. \ref{fig:figure15}, electron density profiles at $t=1$ ps in different cases are given. In general, with the increasing of blow-off plasma density from 0$n_c$ to 1000$n_c$, the electron density in front of the target declines gradually, which has been discussed in the previous section to be important in both the second stage and the third stage. The difference of electron density in front of the target is caused by the electron recirculation inside the target, which is influenced by blow-off plasma density. Due to the presence of the sheath field behind the target, a large part of electrons did not leave from the right boundary directly but return to the left until they came to the target front surface, forming the reflux current and electron recirculation inside the target. Fig. \ref{fig:figure16}(a) and (b) show the energy spectrum of the electrons which travel forwards through the observation plane at x=34 $\mu$m and travel backwards through the observation plane at x=21 $\mu$m respectively. When the density of blow-off plasma increases to 100$n_c$, the number of electrons reflected back by the sheath field behind the target is reduced, which explains why the laser energy absorption rate decreases significantly in collisionless 100$n_c$ case.

The first column of Fig.\ \ref{fig:figure17}\ shows the longitudinal phase plot at $t=1$ ps of different blow-off plasma densities for collisionless cases while the second column for collisional cases. With the increasing of blow-off plasma density, electrons in the forward bunches and the reflux current are less powerful. For collisionless cases, the electron density of blow-off plasma has a significant effect on the electron recirculation. However, when collision is taken into account, the electron recirculation is suppressed and the effect of blow-off plasma density will not be so significant. As shown in Fig. \ref{fig:figure14}, for collisional cases, laser energy absorption rate remains almost the same for different blow-off plasma densities which means that the electron circulation inside the target is negligible in the presence of collision.

To quantitatively analyze the inhibitory effect of collision on reflux current, we make an estimate of the stopping power within the target. The stopping power inside the target consists of ohmic component $\epsilon_o$ and collisional component $\epsilon_c$ \cite{r29}:

\begin{equation}
\begin{aligned}
    \frac{dE}{dx}&=\epsilon_o+\epsilon_c \\
    &=-\frac{e^2q_0}{\sigma m_e c^2(\gamma_0-1)}(\frac{{\gamma_0}^2}{{\gamma_0}^2-1})^{1/2}\frac{1}{\sqrt{\Gamma(E)}}\\
    &-\frac{4\pi e^4n_i}{m_e c^2R}\Gamma(E)[Z_i\Lambda_{fe}+(Z-Z_i)\Lambda_{be}]
\end{aligned}
\end{equation}

Where $E$ is the energy of fast electron, $\gamma=1+E/m_ec^2$ is the relativistic factor for the electron, and $\Gamma(E)=\gamma^2/(\gamma^2-1)$. $E_0$ and $\gamma_0$ are the initial electron energy and the relativistic factor of electron respectively. $q_0$ is the initial energy flux density of fast electrons. $m_e$, $c$ and $n_i$ are the electron rest mass, the speed of light and ion density respectively. $Z_i=40$ is charge state of Au ions and $Z=79$ is the maximum charge state of Au ions. $\Lambda_{fe}$ and $\Lambda_{be}$ are the Coulomb logarithms for fast electron that collides with free and bound electrons, respectively. $\sigma$ is the plasma conductivity while R is the electron-ion scattering factor:

\begin{equation}
    R=[1-\mathrm{exp}(-\frac{\gamma\pi^2}{4Z_i}\frac{[Z_i\Lambda_{fe}+(Z-Z_i)\Lambda_{be}]}{\Lambda_i})]^{1/2}
\end{equation}

Where $\Lambda_i=\mathrm{ln}[2m_ec^3(\gamma-1)(\gamma^2-1)^{1/2}/(Z_ie^2\omega_p\gamma)]$ is the Coulomb logarithm of fast electrons that collide with ions and $\omega_p$ is the plasma frequency.

To simplify the calculation, here we directly give out the approximate formulas for ohmic and collisional stopping power \cite{r29}:

\begin{equation}
    \epsilon_o\:[\mathrm{MeV/\mu m}]\approx0.125\frac{\chi E}{\lambda^2(1+5.5\cdot10^2T^{3/2}Z_i^{-1}\Lambda_{ei}^{-1})}
\end{equation}
\begin{equation}
    \epsilon_c\:[\mathrm{MeV/\mu m}]\approx0.32\cdot10^{-4}\frac{\rho[Z_i\Lambda_{fe}+(Z-Z_i)\Lambda_{be}]}{ZR}
\end{equation}

Here $E$ is electron energy in the unit of MeV, $\chi\approx0.2$ is laser energy absorption rate, $\lambda=1$ is the laser wavelength in the unit of $\mu$m, $T\approx4$ is target temperature in the unit of keV and $\rho=19.32$ is target density in the unit of g/cc. Take a reasonable approximate value of $\Lambda_{ei}\approx10$, $\Lambda_{fe}\approx6$, $\Lambda_{be}\approx2$ and $R\approx0.6$. Then it can be obtained that for a typical electron generated in the third stage with energy 0.1 MeV, the ohmic stopping power $\epsilon_o\approx2\times10^{-4}$ MeV/$\mu$m and the collisional stopping power $\epsilon_c\approx4\times10^{-3}$ MeV/$\mu$m which is an order of magnitude higher than the ohmic stopping power. At this point, the phenomenon observed in Fig. \ref{fig:figure14}(a) can be explained. Fast electrons generated near the target front surface pass through the target and are reflected by the sheath field at the rear surface. The reflected electrons then pass through the target back again to the front surface, forming the electron recirculation. In this process, the distance that electrons travelled is twice the target thickness (20 $\mu$m). For collisional case, the energy loss of these electrons is $\Delta E\approx0.084$ MeV. Therefore the number of low-energy electrons ($E\sim0.1$ MeV) in the recirculation is greatly reduced, resulting in relatively low laser energy absorption rates. In addition, collision can change the direction of electron velocity. As shown in Fig. \ref{fig:figure18}, transverse momentum of electrons in collisional case is significantly larger than that in collisionless case. This effect also inhibits electron recirculation because the longitudinal momentum of the electron is transferred to the transverse.

\maketitle
\section{Summary}
We have described in detail the three-stage model of laser-plasma interaction, introducing the mechanism of fast electron generation in each stage. Using the three-stage model, the roles of ionization, collision, and blow-off plasma in laser target interaction and fast electron generation are illustrated. With the increase of the charge state of Au ions, the laser-plasma interaction transfers to the later stage, resulting in a decrease in laser energy absorption rate. Collision provides a thermal pressure that makes it easier for electrons to escape into the vacuum in front of the target and be accelerated. On the other hand collision increases the stopping power within the target to decelerate the reflux current of electrons. Therefore collision is beneficial for laser energy absorption in the second stage but will be harmful in the third stage. When there is blow-off plasma behind the target, the electron density of blow-off plasma has a significant impact on the laser absorption rate and electron recirculation in collisionless cases while electron recirculation inside the target will be suppressed severely in collisional cases. Therefore the laser energy absorption remains low and almost unchanged for different blow-off plasma densities in collisional cases. The results show that in the presence of collision, the electron circulation inside the target is negligible. Based on the results of this paper, it is recommended to apply a layer of low-Z material on the inner surface of the gold cone to prevent the laser-plasma interaction from entering the third stage. In addition, the thickness of the gold target should be thin to reduce the energy loss and momentum direction change of the fast electron beam during transport in the gold cone.

\maketitle
\section{ACKNOWLEDGMENTS}
This work is supported by the Strategic Priority Research Program of Chinese Academy of Sciences (Grant Nos. XDA25010100 and XDA250050500), National Natural Science Foundation of China (Grants No. 12075204), and Shanghai Municipal Science and Technology Key Project (No. 22JC1401500). D. Wu thanks the sponsorship from Yangyang Development Fund.


\end{document}